\documentclass[aps,floatfix,showpacs,twocolumn,superscriptaddress]{revtex4} 

\usepackage{graphicx} 
\usepackage{dcolumn} 
\usepackage{amsmath}

\usepackage{graphicx}
\usepackage{longtable}

\begin{document}

\title{Flavour symmetry breaking and meson masses}

\author{Mandar S.\ Bhagwat}
\affiliation{Physics Division, Argonne National Laboratory, 
             Argonne, IL 60439-4843, U.S.A.} 

\author{Lei Chang}
\affiliation{Department of Physics, Peking University, Beijing
100871, China}

\author{Yu-Xin Liu}
\affiliation{Department of Physics, Peking University, Beijing 100871, China} \affiliation{The Key Laboratory of Heavy Ion Physics, Ministry of Education,Beijing
100871, China } 
\affiliation{Center of Theoretical Nuclear Physics, National Laboratory of Heavy Ion Accelerator, Lanzhou 730000, China}

\author{Craig D.\ Roberts}
\affiliation{Physics Division, Argonne National Laboratory, 
             Argonne, IL 60439-4843, U.S.A.} 

\author{Peter C.\ Tandy}
\affiliation{Center for Nuclear Research, Department of Physics, Kent State University, Kent, Ohio 44242, U.S.A.}

\date{\today}

\begin{abstract}
The axial-vector Ward-Takahashi identity is used to derive mass formulae for neutral pseudoscalar mesons.  Flavour symmetry breaking entails non-ideal flavour content for these states.  Adding that the $\eta^\prime$ is not a Goldstone mode, exact chiral-limit relations are developed from the identity.  They connect the dressed-quark propagator to the topological susceptibility.  It is confirmed that in the chiral limit the $\eta^\prime$ mass is proportional to the matrix element which connects this state to the vacuum via the topological susceptibility.  The implications of the mass formulae are illustrated using an elementary dynamical model, which includes an \textit{Ansatz} for that part of the Bethe-Salpeter kernel related to the non-Abelian anomaly.  In addition to the current-quark masses, the model involves two parameters, one of which is a mass-scale.  It is employed in an analysis of pseudoscalar- and vector-meson bound-states.  While the effects of $SU(N_f=2)$ and $SU(N_f=3)$ flavour symmetry breaking are emphasised, the five-flavour spectra are described.  Despite its simplicity, the model is elucidative and phenomenologically efficacious; e.g., it predicts $\eta$--$\eta^\prime$ mixing angles of $\sim - 15^\circ$ and $\pi^0$--$\eta$ angles of $\sim 1^\circ$.  
\end{abstract}


\maketitle

\section{Introduction}
Flavour symmetry breaking has long been of interest.  For example, it showed up in the application of current algebra to strong interaction phenomena.  In QCD no two current-quark masses are equal.  Isospin ($SU(2)$-flavour) breaking is determined by the current-mass difference $m_u-m_d$, while $SU(3)$-flavour breaking can be measured via $m_s-(m_u+m_d)/2$.  The $c$- and $b$-quark current-masses are too large for any sensible discussion of larger flavour symmetry groups but the light-quark mass differences also have an impact on the spectrum of hadrons containing a heavy-quark.  

It is of interest to explore and determine the effect of these differences in current-quark mass throughout the hadron spectrum.  This leads one to consider the difference in mass between charged and neutral hadrons.  Part of that splitting is electromagnetic in origin but constraining the strong component is necessary before one can know just how large that electromagnetic contribution might be.  

We focus herein on the strong interaction component alone.  To be specific, we  concentrate on exploring the effect of flavour-symmetry breaking on pseudoscalar- and vector-meson masses.  These sectors are of particular interest because any reliable calculation of pseudoscalar meson masses must involve a consideration of the axial-vector Ward-Takahashi identity.  Moreover, since when viewed simply vector mesons are spin-flip partners of the pseudoscalars, it is natural to examine how the $1^-$--$0^-$ mass splitting evolves with current-quark mass and mass difference.  On the other hand and in addition, one might find that the strong breaking effects in mesons can be interpreted judiciously and used to inform results for baryons.  

In our analysis we employ the Dyson-Schwinger equations (DSEs), pedagogical introductions to which can be found in Refs.\,\cite{Holl:2006ni,Roberts:2007jh}.  The approach is particularly well suited to the study of bound states upon which symmetries and the manner in which they are broken have a heavy impact \cite{Maris:1997hd}.  Within this framework, using the rainbow-ladder truncation -- the lowest-order in a systematic symmetry-preserving truncation scheme \cite{Munczek:1994zz,Bender:1996bb}, effects of flavour-symmetry breaking were reported in Ref.\,\cite{Jain:1993qh}.  Those results can provide a useful comparison with ours.  However, while in one sense a simpler interaction is employed herein, we have a different perspective and will demonstrate effects that arise in proceeding beyond the leading-order truncation.  This is especially noteworthy in connection with neutral pseudoscalar mesons, for which the non-Abelian anomaly plays an important role \cite{Christos:1984tu}.

In Sec.\,\ref{sec:anomaly} we present the $U(N_f)$ axial-vector Ward-Takahashi identity in its general form.  It necessarily includes a contribution from the non-Abelian anomaly, which is explicated.  We elucidate a couple of exact results that follow from the spectral feature that the $\eta^\prime$ mass is much larger than that of other light-quark pseudoscalar mesons.  In Sec.\,\ref{sectionmass} we derive and discuss exact mass formulae for pseudoscalar mesons.  Section~\ref{examples} introduces a model that enables the illustration of implications of these formulae.  It also reports a calculation of the masses of ground state pseudoscalar and vector mesons for $N_f=5$, and covers the phenomena of mixing between the $\pi^0$, $\eta$ and $\eta^\prime$.  The results enable us to provide estimates of the non-electromagnetic component of the neutron-proton mass difference and the masses of hitherto unseen $B^\ast_f$ mesons.
Section~\ref{summary} recapitulates on the main qualitative features emphasised by our study.

\section{Axial-vector Ward-Takahashi Identity}
\label{sec:anomaly}
The axial-vector Ward-Takahashi identity is basic to any study of pseudoscalar mesons.  The impact of this statement of chiral symmetry and the pattern by which it is broken is felt even by heavy-light \cite{Ivanov:1998ms} and heavy-heavy bound-states \cite{Bhagwat:2006xi}.  

The general form of the identity can be expressed \cite{fn:Euclidean}
\begin{eqnarray}
%
%
%
\nonumber P_\mu \Gamma_{5\mu}^a(k;P)& =& {\cal S}^{-1}(k_+) i \gamma_5 {\cal F}^a 
+ i \gamma_5 {\cal F}^a {\cal S}^{-1}(k_-)
\\
 & &  
- 2 i {\cal M}^{ab}\Gamma_5^b(k;P)  - {\cal A}^a(k;P)\,,
\label{avwti}
\end{eqnarray}
where $P=p_1-p_2$ is the total and $k$ the relative momentum between the amputated quark legs \cite{fn:momentum}.  Eq.\,(\ref{avwti}) is fully renormalised and it is important that the product ${\cal M}^{ab}\Gamma_5^b$ does not mix with other operators under renormalisation.

In Eq.\,(\ref{avwti}), $\{{\cal F}^a | \, a=0,\ldots,N_f^2-1\}$ are the generators of $U(N_f)$ in the fundamental representation, orthonormalised according to tr${\cal F}^a {\cal F}^b= \frac{1}{2}\delta^{ab}$.  The dressed-quark propagator ${\cal S}=\,$diag$[S_u,S_d,S_s,S_c,S_b,\ldots]$ is matrix-valued with nonzero entries that can be expressed in various equivalent forms, e.g., 
\begin{equation}
\label{Spform}
S(k) = \frac{1}{i\gamma\cdot k A(k^2) + B(k^2)} 
= \frac{Z(k^2)}{i \gamma\cdot k + M(k^2)}.
\end{equation}
The propagator is determined by the gap equation
\begin{eqnarray}
\lefteqn{{\cal S}(p)^{-1} =  Z_2 \,(i\gamma\cdot p + {\cal M}^{\rm bm}) + \Sigma(p)\,,} \label{gendse} \\
\Sigma(p) & = & Z_1 \int^\Lambda_q\! g^2 D_{\mu\nu}(p-q) \frac{\lambda^a}{2}\gamma_\mu {\cal S}(q) \Gamma^a_\nu(q,p) , \label{gensigma}
\end{eqnarray}
wherein: $\int^\Lambda_q$ represents a Poincar\'e invariant regularisation of the integral, with $\Lambda$ the regularisation mass-scale \cite{Maris:1997hd}; $D_{\mu\nu}$ is the dressed-gluon propagator; $\Gamma_\nu(q,p)$ is the dressed-quark-gluon vertex; and ${\cal M}^{\rm bm} = {\rm diag}[m_u^{\rm bm},m_d^{\rm bm},m_s^{\rm bm},m_c^{\rm bm},m_b^{\rm bm},\ldots]$ is the matrix of $\Lambda$-dependent current-quark bare masses.  The quark-gluon-vertex and quark wave function renormalisation constants, $Z_{1,2}(\zeta^2,\Lambda^2)$, depend on the gauge parameter, the renormalisation point, $\zeta$, and the regularisation mass-scale.  The gap equation is completed by the renormalisation condition 
\begin{equation}
\label{renormS} \left.{\cal S}(p)^{-1}\right|_{p^2=\zeta^2} = i\gamma\cdot p +
{\cal M}(\zeta)\,,
\end{equation}
where ${\cal M}(\zeta)$ is the matrix of renormalised (running) current-quark masses whose nonzero entries obey
\begin{equation}
Z_2(\zeta^2,\Lambda^2) \, m^{\rm bm}(\Lambda) = Z_4(\zeta^2,\Lambda^2) \, m(\zeta)\,,
\end{equation}
with $Z_4$ the Lagrangian mass renormalisation constant.  In Eq.\,(\ref{avwti}) we have defined 
\begin{equation}
{\cal M}^{ab} = {\rm tr}_F \left[ \{ {\cal F}^a , {\cal M} \} {\cal F}^b \right],
\end{equation}
where the trace is over flavour indices.

The inhomogeneous axial-vector vertex in Eq.\,(\ref{avwti}) satisfies 
\begin{eqnarray}
\nonumber
\lefteqn{\left[\Gamma^a_{5\mu}(k;P)\right]_{tu}
=  Z_2 \left[\gamma_5\gamma_\mu {\cal F}^a \right]_{tu}}\\
 &+& \int^\Lambda_q
[{\cal S}(q_+) \Gamma^a_{5\mu}(q;P) {\cal S}(q_-)]_{sr} K_{tu}^{rs}(q,k;P)\,,
\label{avbse}
\end{eqnarray}
where $P=p_1-p_2=q_1-q_2$ and $r$,\ldots,\,$u$ represent colour, Dirac and flavour indices.  The pseudoscalar vertex $\Gamma_{5}^a$ satisfies an analogous equation driven by the inhomogeneity $Z_4 \gamma_5 {\cal F}^a$.  

The final term in the last line of Eq.\,(\ref{avwti}) expresses the non-Abelian axial anomaly.  It can be written
\begin{equation}
\label{amputate}
{\cal A}^a(k;P) =  {\cal S}^{-1}(k_+) \,\delta^{a0}\, {\cal A}_U(k;P) {\cal S}^{-1}(k_-)\,,
\end{equation}
with
\begin{equation}
{\cal A}_U(k;P) = \!\!  \int\!\! d^4xd^4y\, e^{i(k_+\cdot x - k_- \cdot y)} N_f \left\langle  {\cal F}^0\,q(x)  \, {\cal Q}(0) \,   \bar q(y) 
\right\rangle. \label{AU}
\end{equation}
Here the matrix element represents an operator expectation value in full QCD; the operation in Eq.\,(\ref{amputate}) amputates the external quark lines; and 
\begin{equation}
{\cal Q}(x) = i \frac{\alpha_s }{4 \pi} {\rm tr}_{C}\left[ \epsilon_{\mu\nu\rho\sigma} F_{\mu\nu} F_{\rho\sigma}(x)\right]  \label{topQ}\\
= \partial_\mu K_\mu(x)
\end{equation}
is the topological charge density operator, where the trace is over colour indices and $F_{\mu\nu}=\frac{1}{2}\lambda^a F_{\mu\nu}^a$ is the matrix-valued gluon field strength tensor.  It is plain and important that only ${\cal A}^{a=0}$ is nonzero.  NB.\ While ${\cal Q}(x)$ is gauge invariant, the associated Chern-Simons current, $K_\mu$, is not.  

If one imagines there are $N_f$ massless quarks, then dynamical chiral symmetry breaking (DCSB) is a necessary and sufficient condition for the $a\neq 0$ components of Eq.\,(\ref{avwti}) to guarantee the existence of $N_f^2-1$ massless bound-states of a dressed-quark and -antiquark \cite{Maris:1997hd}.  

However, owing to Eq.\,(\ref{amputate}), $a=0$ in Eq.\,(\ref{avwti}) requires special consideration.  One case is easily covered; viz., it is clear that if ${\cal A}^{0} \equiv 0$, then the $a=0$ component of Eq.\,(\ref{avwti}) is no different to the others and there is an additional massless bound-state in the chiral limit.  

On the other hand, the large disparity between the mass of the $\eta^\prime$-meson and the octet pseudoscalars suggests that ${\cal A}^{0} \neq 0$ in real-world QCD.  Let's consider this possibility and proceed by allowing that $\Gamma_{5\mu}^0$ might possess a longitudinal massless bound-state pole.  In this case one can write
\begin{eqnarray}
\nonumber \lefteqn{\left.\Gamma_{5\mu}^0(k;P)\right|_{P^2 \approx 0}=
 r_A^0\frac{P_\mu}{P^2 } \Gamma_{BS}(k;P)  }\\
\nonumber
& &+ {\cal F}^0 \gamma_5 \left[ \gamma_\mu F_R^0(k;P)+ \gamma\cdot k k\cdot P G_R^0(k;P) \right. \\
&& \left.  + \sigma_{\mu\nu} k_\mu P_\nu H_R^0(k;P)
 + \tilde\Gamma_{5 \mu}^{0}(k;P) \right]\,,
\label{genavv0} 
\end{eqnarray}
where $F_R^0$, $G_R^0$, $H_R^0$ and $\Gamma_{5 \mu}^{0}(k;P)$ are regular as $P^2\to 0$, $P_\mu \tilde\Gamma_{5 \mu}^{0} \sim $O$(P^2)$, $\Gamma_{BS}(k;P)$ is the possible bound-state's canonically normalised Bethe-Salpeter amplitude and $r_A^0$ is its residue.  The amplitude takes the general form \cite{Llewellyn-Smith:1969az}
\begin{eqnarray}
\nonumber \lefteqn{\Gamma_{BS}(k;P) = 2{\cal F}^0 \gamma_5 \left[ i E_{BS}(k;P) + \gamma\cdot P F_{BS}(k;P) \right.} \\
&& \left. +\gamma\cdot k k\cdot P G_{BS}(k;P) + \sigma_{\mu\nu} k_\mu P_\nu H_{BS}(k;P)\right].
\end{eqnarray}

Since in these circumstances one can write
\begin{eqnarray}
\nonumber \lefteqn{{\cal A}^0(k;P) = {\cal F}^0\gamma_5 \left[ i {\cal E}_{\cal A}(k;P) + \gamma\cdot P {\cal F}_{\cal A}(k;P) \right.} \\
&& \left. +\gamma\cdot k k\cdot P {\cal G}_{\cal A}(k;P) + \sigma_{\mu\nu} k_\mu P_\nu {\cal H}_{\cal A}(k;P)\right],
\end{eqnarray}
then the Goldberger-Treiman relations of Ref.\,\cite{Maris:1997hd} become
\begin{eqnarray}
\label{ewti}
2 r_A^0 E_{BS}(k;0) &= & 2 B_{0}(k^2) - {\cal E}_{\cal A}(k;0),\\
\label{fwti}
F_R^0(k;0) + 2 r_A^0 F_{BS}(k;0) & = & A_{0}(k^2) - {\cal F}_{\cal A}(k;0),\\
G_R^0(k;0) + 2 r_A^0 G_{BS}(k;0) & = & 2 A^\prime_{0}(k^2) - {\cal G}_{\cal A}(k;0),\\
\label{hwti}
H_R^0(k;0) + 2 r_A^0 H_{BS}(k;0) & = & - {\cal H}_{\cal A}(k;0),
\end{eqnarray}
where $A_0$, $B_0$ characterise the gap equation's chiral limit solution.  NB.\ A massless pole in ${\cal A}^0(k;P)$ is incompatible with Eq.\,(\ref{avwti}).

It now plain that if 
\begin{equation}
\label{calEB}
{\cal E}_{\cal A}(k;0) = 2 B_{0}(k^2) \,,
\end{equation}
then $r_A^0 E_{BS}(k;0) \equiv 0$.  This being true, then the homogeneous Bethe-Salpeter equation (BSE) also produces $r_A^0 F_{BS}(k;0)\equiv 0 \equiv r_A^0 G_{BS}(k;0) \equiv 2 r_A^0 H_{BS}(k;0)$.  Hence, Eq.\,(\ref{calEB}) guarantees that $\Gamma_{5\mu}^0$ \emph{cannot} posses a massless pole.  The converse is also true; namely, the absence of such a pole requires Eq.\,(\ref{calEB}).  It is noteworthy that in the neighbourhood of $P^2=0$, Eqs.\,(\ref{fwti}) -- (\ref{hwti}) thus provide pointwise relations between ${\cal A}^0(k;P)$, the dressed-quark propagator and the regular part $\Gamma_{5\mu}^0(k;P)$.

Equation~(\ref{calEB}) is a necessary and sufficient condition for the absence of a massless bound-state pole in $\Gamma_{5\mu}^0$.  We are discussing the chiral limit, in which case $B_{0}(k^2) \neq 0 $ if, and only if, chiral symmetry is dynamically broken.   Hence, the absence of an additional massless bound-state is only assured through the existence of an intimate connection between DCSB and an expectation value of the topological charge density.  

This noteworthy connection is further highlighted by the following result, obtained through a few straightforward manipulations of Eqs.\,(\ref{avwti}), (\ref{amputate}) and (\ref{AU}):
\begin{eqnarray}
\langle \bar q q \rangle_\zeta^0 & = & -\lim_{\Lambda\to \infty} 
Z_4(\zeta^2,\Lambda^2)\, {\rm tr}_{\rm CD}\int^\Lambda_q\!
S^{0}(q,\zeta)  \\
& = & 
\mbox{\footnotesize $\displaystyle \frac{N_f}{2}$} \int d^4 x\, \langle \bar q(x) i\gamma_5  q(x) {\cal Q}(0)\rangle^0,
\end{eqnarray}
where here the superscript ``0'' indicates that the quantity is calculated in the chiral limit.  The absence of a Goldstone boson in the $a=0$ channel is only guaranteed if this explicit identity between the chiral-limit vacuum quark condensate and the vacuum polarisation generated by the topological charge density is satisfied.  

\section{Mass Formulae}
\label{sectionmass}
A wide range of additional observations are possible, some of which are canvassed in Ref.\,\cite{Christos:1984tu}.  Herein we will derive those that are especially relevant in the context of this work.  

Equation\,(\ref{avwti}) is an identity that connects two and three point functions in QCD.  It applies at all values of the total momentum $P$, in particular, at the location of bound-state poles.  To exploit this we extend Eq.\,(\ref{genavv0}) and observe that in the neighbourhood of such a pseudoscalar pole, whether massless or massive,
\begin{eqnarray}
\nonumber \left.\Gamma_{5\mu}^a(p_1,p_2)\right|_{P^2+m_{\pi_i}^2 \approx 0}&=&   \frac{f_{\pi_i}^a \, P_\mu}{P^2 + 
m_{\pi_i}^2} \Gamma_{\pi_i}(k;P)\\
& & + \; \Gamma_{5 \mu}^{a\,{\rm reg}}(p_1,p_2) \,, \label{genavv} \\
\nonumber 
\left. i\Gamma_{5 }^a(p_1,p_2)\right|_{P^2+m_{\pi_i}^2 \approx 0}
&=&   \frac{\rho_{\pi_i}^a(\zeta) }{P^2 + 
m_{\pi_n}^2} \Gamma_{\pi_i}(k;P)\\
& & + \; i\Gamma_{5 }^{a\,{\rm reg}}(k;P) \,; \label{genpv} 
\end{eqnarray}
viz., each vertex in Eq.\,(\ref{avwti}) is expressed as a simple pole plus terms regular in the neighbourhood of this pole, with $\Gamma_{\pi_i}(k;P)$ representing the bound-state's canonically normalised Bethe-Salpeter amplitude \cite{Llewellyn-Smith:1969az}, where $i=0$ labels the lightest pseudoscalar bound-state, $i=1$, the next lightest, and so on.  In Eqs.\,(\ref{genavv}) and (\ref{genpv})
\begin{eqnarray} 
\label{fpia} f_{\pi_i}^a \,  P_\mu &=& Z_2\,{\rm tr} \int^\Lambda_q 
{\cal F}^a \gamma_5\gamma_\mu\, \chi_{\pi_i}(q;P) \,, \\
\label{cpres} i  \rho_{\pi_i}^a\!(\zeta)  &=& Z_4\,{\rm tr} 
\int^\Lambda_q {\cal F}^a \gamma_5 \, \chi_{\pi_i}(q;P)\,,
\end{eqnarray} 
where $\chi_{\pi_i}(k;P) = {\cal S}(k_+) \Gamma_{\pi_i}(k;P) {\cal S}(k_-)$, $k_\pm = q\pm P/2$.  These residues are gauge invariant and cutoff independent.  NB.\ The nature of $r_A^0$ in Eq.\,(\ref{genavv0}) is now clear.

While there is certainly no bound-state pole in the inverse of the dressed-quark propagator, the opposite can be true of the term associated with the topological susceptibility; namely, we must consider 
\begin{eqnarray}
\nonumber 
\left. {\cal A}^0(p_1,p_2) = \right|_{P^2+m_{\pi_i}^2 \approx 0} &=&   \frac{ n_{\pi_i}}{P^2 + 
m_{\pi_i}^2} \Gamma_{\pi_i}(k;P)\\
& & + \; {\cal A}^{0\,{\rm reg}}(p_1,p_2), \label{genA}
\end{eqnarray}
where 
\begin{equation}
n_{\pi_i} = \mbox{\footnotesize $\displaystyle \sqrt{\frac{N_f}{2}}$} \, \nu_{\pi_i} \,, \; \nu_{\pi_i}= \langle 0 | {\cal Q} | \pi_i\rangle \,.
\end{equation}

Using Eqs.\,(\ref{genavv}), (\ref{genpv}) and (\ref{genA}) in the axial-vector Ward-Takahashi identity, we arrive at a mass formula for pseudoscalar mesons:
\begin{equation}
\label{newmass}
m_{\pi_i}^2 f^a_{\pi_i} = 2\,{\cal M}^{ab} \rho_{\pi_i}^b + \delta^{a0}n_{\pi_i}\,.
\end{equation} 
It is valid for current-quark masses of any magnitude.

For nondiagonal mesons this is naturally the same formula as derived in Refs.\,\cite{Maris:1997hd,Maris:1997tm}.  Allowing for the fact that the Standard Model requires observable particles to be eigenstates of the electric charge, it yields, e.g., with ${\cal F}^{K^+}={\cal F}^4 - i {\cal F}^5$
\begin{equation}
m_{K^+}^2 f_{K^+} = [m_u(\zeta) + m_s(\zeta)] \, \rho_{K^+}(\zeta)\,,
\end{equation}
and similarly, 
\begin{equation}
m_{D^+}^2 f_{D^+} = [m_d(\zeta) + m_c(\zeta)] \, \rho_{D^+}(\zeta)\,.
\end{equation}
(NB.\ With our normalisation, $f_{K^+}= 113\,$MeV experimentally.)  Again, these formulae are valid for arbitrarily large, or small, current-quark masses.  The Gell-Mann--Oakes--Renner relation is a small quark mass corollary \cite{Maris:1997hd,Maris:1997tm} and aspects of their implications for mesons containing a heavy-quark, or two, are detailed in Refs.\,\cite{Ivanov:1998ms,Bhagwat:2006xi}.

A novelty of Eq.\,(\ref{newmass}) is its validity for charge neutral mesons.  For example, in the case $N_f=3$ one derives for the neutral pion
\begin{equation}
m_{\pi^0}^2 \left[
\begin{array}{c}
f_{\pi^0}^3 \\[1ex] f_{\pi^0}^8 \\[1ex] f_{\pi^0}^0 
\end{array} \right] = 
\left[ \begin{array}{c}
0 \\ 0 \\ n_{\pi^0} \end{array}\right] + \bigg[ M_{3\times 3} \bigg] 
\left[\begin{array}{c}
\rho_{\pi^0}^3 \\[1ex] \rho_{\pi^0}^8 \\[1ex] \rho_{\pi^0}^0 
\end{array}\right],
\label{MassFormpi0}
\end{equation}
where 
\begin{equation}
\bigg[ M_{3\times 3} \bigg]\!  = \!\left[
\begin{array}{ccc}
m_{110} & \sqrt{\frac{1}{3}} \, m_{1-10} & \sqrt{\frac{2}{3}}\, m_{1-10}\\
\sqrt{\frac{1}{3}}\, m_{1-10}& \frac{1}{3} \, m_{114} & \sqrt{\frac{2}{9}}\, m_{11-2}\\
\sqrt{\frac{2}{3}} \, m_{1-10} & \sqrt{\frac{2}{9}} \, m_{11-2} & \frac{2}{3}\, m_{111}
\end{array}\right],
\end{equation}
with  $m_{\alpha\beta\gamma}= \alpha \,m_u + \beta\, m_d + \gamma \,m_s$.  

In the isospin symmetric case; i.e., $m_u=m_d$, $M_{3\times 3}$ exhibits no mixing between ${\cal F}^3$ and ${\cal F}^{0,8}$.  This signals that the flavour content of the $\pi^0$ is described solely by ${\cal F}^3$.  Therefore Eqs.\,(\ref{fpia}) and (\ref{cpres}) give $f_{\pi^0}^8= 0= f_{\pi^0}^0$, $\rho_{\pi^0}^8= 0= \rho_{\pi^0}^0$, and Eq.\,(\ref{genA}) yields $\nu_{\pi^0}=0$.  Hence, in this instance the complete content of Eq.\,(\ref{MassFormpi0}) is
\begin{equation}
m_{\pi^0}^2 f_{\pi^0}^3 = [m_u(\zeta) + m_d(\zeta)] \, \rho_{\pi^0}^3(\zeta).
\end{equation}
This is not true, however, for $m_u \neq m_d$, as we shall subsequently illustrate.

For $N_f=3$ one also obtains
\begin{eqnarray}
m_{\eta}^2 \left[
\begin{array}{c}
f_{\eta}^3 \\[1ex] f_{\eta}^8 \\[1ex] f_{\eta}^0 
\end{array} \right] & = & 
\left[ \begin{array}{c}
0 \\ 0 \\ n_{\eta} \end{array}\right] + \bigg[ M_{3\times 3} \bigg] 
\left[\begin{array}{c}
\rho_{\eta}^3 \\[1ex] \rho_{\eta}^8 \\[1ex] \rho_{\eta}^0 
\end{array}\right], \label{MassFormeta0} \\
m_{\eta^\prime}^2 \left[
\begin{array}{c}
f_{\eta^\prime}^3 \\[1ex] f_{\eta^\prime}^8 \\[1ex] f_{\eta^\prime}^0 
\end{array} \right] & = & 
\left[ \begin{array}{c}
0 \\ 0 \\ n_{\eta^\prime} \end{array}\right] + \bigg[ M_{3\times 3} \bigg] 
\left[\begin{array}{c}
\rho_{\eta^\prime}^3 \\[1ex] \rho_{\eta^\prime}^8 \\[1ex] \rho_{\eta^\prime}^0 
\end{array}\right]. \label{MassFormetap} 
\end{eqnarray}
Naturally, on the domain in which an expansion in current-quark mass is valid; viz., $m(1\,{\rm GeV})\lesssim 50\,$MeV \cite{Chang:2006bm}, Eqs.\,(\ref{MassFormpi0}), (\ref{MassFormeta0}) and (\ref{MassFormetap}) reproduce current algebra results \cite{Gasser:1982ap}.  

Of importance is a prediction of the manner by which the $\eta^\prime$ is split from the octet pseudoscalars by an amount that depends on QCD's topological susceptibility.  This is most easily illustrated by considering the $U(N_f)$ limit, in which all current-quark masses assume the single value $m(\zeta)$.  In this case the complete content of Eq.\,(\ref{MassFormetap}) is the statement
\begin{equation}
\label{etapchiral}
m_{\eta^\prime}^2 f_{\eta^\prime}^0 = n_{\eta^\prime} + 2 m(\zeta)\rho_{\eta^\prime}^0(\zeta) \,.
\end{equation}
Plainly, the $\eta^\prime$ is split from the Goldstone modes so long as $n_{\eta^\prime} \neq 0$ \cite{fn:ghost}.  Numerical simulations of lattice-regularised QCD have confirmed the relationship reproduced here \cite{Bardeen:2000cz,Ahmad:2005dr}.  

It is argued \cite{Witten:1979vv,Veneziano:1979ec} that in QCD 
\begin{equation}
n_{\eta^\prime} \sim \frac{1}{\sqrt{N_c}}\,,
\end{equation}
and it can be seen to follow from the gap equation, the homogeneous BSE and Eqs.\,(\ref{fpia}), (\ref{cpres}) that 
\begin{equation}
f_{\eta^\prime}^0 \sim \sqrt{N_c} \sim \rho_{\eta^\prime}^0(\zeta)\,.
\end{equation}
One thus obtains 
\begin{equation}
m_{\eta^\prime}^2 =  \frac{n_{\eta^\prime}}{f_{\eta^\prime}^0} + 2 m(\zeta) \frac{\rho_{\eta^\prime}^0(\zeta)}{f_{\eta^\prime}^0} \,.
\end{equation}
The first term vanishes in the limit $N_c\to \infty$ while the second remains finite.  Subsequently taking the chiral limit, the $\eta^\prime$ mass approaches zero in the manner characteristic of all Goldstone modes.  (NB.\ One must take the limit $N_c\to \infty$ before the chiral limit because the procedures do not commute \cite{Narayanan:2004cp}.)  These results are realised in the effective Lagrangian of Ref.\,\cite{Di Vecchia:1979bf} in a fashion that is consistent with all the constraints of the anomalous Ward identity \cite{fn:thooft}.

\section{Meson Masses: Exemplifying Effects of Mixing}
\label{examples}
\subsection{Model defined}
Implications of the exact results presented above can be illustrated and further elucidated by way of a simple kernel for the gap and BSEs.  We write
\begin{equation}
\label{KLA}
K = K_{L} + K_A\,,
\end{equation}
where $K_L$ is the leading order in the systematic and symmetry preserving truncation explained in Refs.\,\cite{Munczek:1994zz,Bender:1996bb}; namely, a dressed-ladder interaction:
\begin{eqnarray}
\nonumber \lefteqn{
(K_L)^{tu}_{rs}(q,p;P) =  }\\
&& \!\!\! - \,{\cal G}((p-q)^2) \, D_{\mu\nu}^{\rm free}(p-q)
 \,\left[\gamma_\mu \frac{\lambda^a}{2}\right]_{ts} \, \left[\gamma_\nu \frac{\lambda^a}{2}\right]_{ru} \!\!\!, \label{ladderK}
\end{eqnarray}
wherein $D_{\mu\nu}^{\rm free}(k)$ is the free gauge boson propagator and ${\cal G}(k^2)$ represents an effective coupling.  For the latter we use the simple model introduced in Ref.\,\cite{Munczek:1983dx}
\begin{equation}
\label{Gk}
{\cal G}(k^2) = (2\pi)^4 {\cal G}^2 k^2 \delta^4(k)
\end{equation}
with ${\cal G}$ a constant that sets the mass-scale.  The model is ultraviolet-finite and hence one can remove the regularisation mass-scale to infinity and set the renormalisation constants to one.  The infrared enhancement exhibited by Eq.\,(\ref{Gk}) is sufficient to provide for confinement and DCSB, as explained, e.g., in Sec.\,2.2 of Ref.\,\cite{Roberts:2007jh}.  Moreover, in practice it has many features in common with a class or renormalisation-group-improved effective-interactions; and its distinctive momentum-dependence works to advantage in reducing integral equations to algebraic equations that preserve the character of the original.  There is a drawback: the simple momentum dependence can lead to some model-dependent artefacts, but they are easily identified and hence not generally cause for serious concern.  

One weakness hampers us, however.  The model generates Bethe-Salpeter amplitudes whose dependence on the relative momentum, $k$, is unrealistic.  In an internally consistent definition it is described by $\delta^4(k)$.  Hence one cannot obtain values for the independent overlaps $f_{\pi_i}^a$ and $\rho_{\pi_i}^a$ in Eq.\,(\ref{newmass}) and so the mass formulae cannot be directly verified.  Nevertheless, we will see their imprint in the calculated results for meson masses.  Verification of Eq.\,(\ref{newmass}) is possible with the interactions employed, e.g., in Refs.\,\cite{Jain:1993qh,Maris:1997tm,Maris:1999nt}.  Indeed, that has already been done for channels in which the rainbow-ladder truncation is a good approximation, such as charged pseudoscalar mesons and neutral heavy-heavy pseudoscalar mesons \cite{Maris:1997tm,Bhagwat:2006xi}.

\begin{figure}[t]
\vspace*{-30ex}

\centerline{
\includegraphics[clip,width=0.5\textwidth]{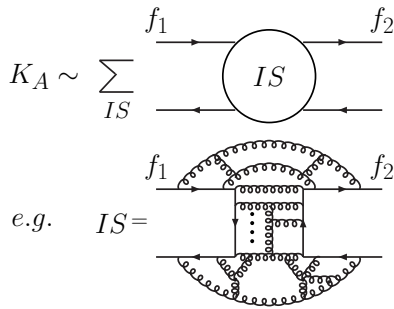}}
\vspace*{-31ex}

\caption{\label{etaglue} An illustration of the nature of $K_A$; viz., the contribution to the Bethe-Salpeter kernel associated with the non-Abelian anomaly.  All terms have the ``hairpin'' structure illustrated in the lower panel.  No finite sum of such intermediate states is sufficient.  Straight lines denote quarks, with $f_1$ and $f_2$ independent, and springs denote gluons.}
\end{figure}

In Eq.\,(\ref{KLA}), $K_A$ is a novel addition that we use to model effects owing to the non-Abelian anomaly.  Its inclusion takes us beyond ladder-truncation and is thus an expedient which is a dynamical extension of that employed in Ref.\,\cite{Klabucar:1997zi}.  It can be argued from Eqs.\,(\ref{AU}) and (\ref{topQ}) that an anomaly-related contribution to a meson's Bethe-Salpeter kernel cannot contain external quark or antiquark lines that are connected to the incoming lines: purely gluonic configurations must mediate, as illustrated in Fig.\,\ref{etaglue}.  Furthermore, it is straightforward to see that no finite sum of gluon exchanges can serve this purpose.  Indeed, consider any one such single contribution in the chiral limit.  It will be proportional to the total momentum and hence vanish for $P=0$, in conflict with Eq.\,(\ref{etapchiral}).  This lies behind the need for something like the Kogut-Susskind \emph{ghost} \cite{fn:ghost}.  (NB.\, The resummed kernels explored in Refs.\,\cite{Bender:2002as,Bhagwat:2004hn,Matevosyan:2006bk} do not resolve such vacuum polarisation diagrams \cite{Bender:1996bb} and thus cannot generate $K_A\neq 0$.)

As in Ref.\,\cite{Holl:2004un}, with these observations in mind we employ
\begin{eqnarray}
\nonumber 
\lefteqn{(K_A)_{rs}^{tu}(q,p;P)  } \\
\nonumber
&=& -\xi((q-p)^2) \left\{\cos^2\theta_\xi \,[\varsigma\gamma_5]_{rs} [\varsigma\gamma_5]_{tu} \right.\\
&&+\left. \sin^2\theta_\xi \,[\varsigma\gamma\cdot P\gamma_5]_{rs} [\varsigma\gamma\cdot P\gamma_5]_{tu}\right\}, \label{defKA}\\
\label{defxi}
\xi(k^2)& =&  (2\pi)^4\, \xi \, \delta^4(k)\,,
\end{eqnarray}
where $\xi$ is a dimensionless coupling strength.  In principle, $\xi(k^2)$ would also depend on the total momentum but for simplicity we ignore that herein.  In proposing Eq.\,(\ref{defKA}) we have also used the fact that Eq.\,(\ref{Gk}) only supports a pseudoscalar Bethe-Salpeter amplitude of the form 
\begin{equation}
\label{BSamp}
\Gamma_{\pi_i}(P) = 2\, {\cal F}^{\pi_i}\gamma_5 \left[ ig_1^{\pi_i} + \gamma\cdot P g_2^{\pi_i}\right]\,;
\end{equation}
namely, as described above, the interaction requires the constituents' relative momentum to vanish.  The angle $\theta_\xi$ controls the relative magnitude of the two possible contributions to the kernel.

The remaining piece of Eq.\,(\ref{defKA}) is the flavour matrix
\begin{eqnarray}
\varsigma & = & {\rm diag} [\frac{1}{M^D_u},\frac{1}{M^D_d},\frac{1}{M^D_s},\ldots ]\,,\\
M^D_f & = & M_f(s=0) \,, \label{MQ}
\end{eqnarray}
where $M_f(s)$ is the mass function for a quark of flavour $f$ [see Eq.\,(\ref{Spform})].  Equation\,(\ref{MQ}) defines a dynamical constituent-quark mass.  It differs from the Euclidean constituent-quark mass (e.g., Ref.\,\cite{Flambaum:2005kc}), but is easier to calculate and is likewise a renormalisation point invariant in QCD.  This term introduces a nonperturbative mass-dependence, which models that arising from the dressed-quark lines that complete a ``U-turn'' in the so-called hairpin diagram in Fig.\,\ref{etaglue}.

\subsection{Parameters fixed}
The model has three parameters in addition to which there are $N_f$ current-quark masses.  We determine the current-quark masses and ${\cal G}$ in Eq.\,(\ref{Gk}) by applying the model to charged-pseudoscalar and vector meson ground-states.  Since $K_A$ doesn't contribute in these channels, this corresponds to a rainbow-ladder treatment of those states, which is plausibly accurate to $\lesssim 10$\% \cite{Matevosyan:2006bk} for light-quark mesons and becomes precise for heavy-heavy systems \cite{Bhagwat:2004hn}.

The rainbow gap equation is obtained from Eq.\,(\ref{gendse}) with
\begin{equation}
\label{SigmaRL}
\Sigma_{tu}(p) = -\int_q \, (K_L)_{rs}^{tu}(q,p;P)\, {\cal S}_{sr}(q)\,.
\end{equation}
(NB.\ It is plain upon insertion that $K_A$ defined in Eq.\,(\ref{defKA}) does not modify Eq.\,(\ref{SigmaRL}).)  The gap equation is solved to obtain the matrix dressed-quark propagator, which is then used to complete the homogeneous BSE:
\begin{equation}
\label{LadderBSE}
\Gamma_{H}(k;P) = \int_q
[{\cal S}(q_+) \Gamma_{H}(q_+,q_-) {\cal S}(q_-)]_{sr} (K_L)_{tu}^{rs}(q,k;P)\,.
\end{equation}

Equation\,(\ref{LadderBSE}) can viewed as defining an eigenvalue problem.  There can be and is no mixing between charged and neutral mesons so the eigenvector in the case of charged pseudoscalars can be written
\begin{equation}
\label{BSampG}
\Gamma_{H_5}(P) = \sum_{d=1,2,4,5,6,7,\ldots} \, 2\,{\cal F}^a \gamma_5 \left[ ip_1^{a} + \gamma\cdot P p_2^{a}\right],
\end{equation}
with the index selecting all $N_f (N_f-1)$ nondiagonal generators of $SU(N_f)$.  One inserts Eq.\,(\ref{BSampG}) into Eq.\,(\ref{LadderBSE}) and evaluates the spinor trace to arrive at an equation with the structure
\begin{equation}
\label{eigenproblem}
\vec{p} = [K_{H_5}] \vec{p}\,,
\end{equation}
with $\vec{p} = {\rm column}[p_1^{1},p_2^{1},p_1^{2},p_2^{2},\ldots]$.  (The procedure is made explicit, e.g., in Ref.\,\cite{Bender:2002as}.)  The matrix $K_{H_5}$ has $N_f (N_f-1)$ eigenvalues, $\{\lambda^i_{H_5}\}$, and eigenvectors, each of which  depend on the value of $P^2$.  One has a solution of the homogeneous BSE when one of those eigenvalues acquires the value ``one'' and the mass of the associated bound-state is the value of $P^2$ for which this occurs; viz., 
\begin{equation}
\label{eigenvalueP}
(m_{H_5}^i)^2 = \{-P^2 \, | \, \lambda^i_{H_5}(P^2) = 1\}\,.
\end{equation}
At this point the related eigenvector is that meson's Bethe-Salpeter amplitude.  If $n$ eigenvectors assume the value ``one'' at the same value of $P^2$, then one has an $n$-fold degeneracy. 

With the interaction in Eq.\,(\ref{Gk}), the eigenvector associated with the vector mesons of $U(N_f)$ has the general form
\begin{equation}
\label{BSampV}
\Gamma_{H_V}^\lambda(P) = \sum_{a=1}^{N_f} \, 2\,{\cal F}^a \left[ \gamma\cdot\epsilon^\lambda \, v_1^{a} + \sigma_{\mu\nu} \epsilon_\mu^\lambda P_\nu \,  v_2^{a}\right],
\end{equation}
where $\{\epsilon_\mu^\lambda(P) | \lambda=-1,0,1\}$ is the polarisation four-vector
\begin{equation}
P\cdot \epsilon^\lambda(P) = 0\,, \; \forall \lambda\,;\; \epsilon^\lambda(P)\cdot \epsilon^{\lambda^\prime}(P) = \delta^{\lambda \lambda^\prime}.
\end{equation}
In ladder truncation, $v_2\equiv 0$.  From this point the solution of Eq.\,(\ref{LadderBSE}) proceeds as described above for charged pseudoscalars.   

We are now in a position to fix the reference values of the mass-scale parameter and the current-quark masses.  In the absence of electromagnetism we fix the values of ${\cal G}$ and the sum $(m_u+m_d)=:2 \bar m$ so as to obtain the experimental values of the ratio $m_{\pi^0}/m_{\rho^0}$ and $m_{\rho^0}$.  Moreover, the pseudoscalar variant of Eq.\,(\ref{LadderBSE}) produces degenerate bound-states associated with ${\cal F}^1$ \& ${\cal F}^2$ -- $\pi^\pm$, two others associated with ${\cal F}^4$ \& ${\cal F}^5$ -- $K^\pm$, another two with ${\cal F}^6$ \& ${\cal F}^7$ -- $K^0$, $\overline{K}^0$, etc.  This is similarly true of the vector equation: $\rho^\pm$, $K^{\ast \pm}$, etc.  We choose to fix the mass difference $(m_u-m_d)$ by requiring 
\begin{equation}
[m_{K^{\ast 0}} - m_{K^{\ast +}}]_{f}= 5.42\,{\rm MeV},
\end{equation}
which is the weighted average of the isospin-only differences estimated in Ref.\,\cite{Schechter:1992iz}.  The rainbow-ladder truncation produces pure $\bar s s$, $\bar c c$ and $\bar b b$ vector mesons and $m_{s,c,b}$ are set by identifying these states with the observed $\phi$-, $J/\psi$- and $\Upsilon$-mesons.   This procedure yields the following values
\begin{eqnarray}
&& {\cal G} =   0.537 \, {\rm GeV}\,, \label{Gval}\\
\nonumber 
&& \begin{array}{lllll}
 m_u & m_d & m_s & m_c & m_b \\\hline
 0.0140 \,{\cal G} & 0.0271\, {\cal G} & 0.323\, {\cal G} & 2.55\, {\cal G} & 8.67\, {\cal G}\\
7.51\,{\rm MeV} & 14.6\,{\rm MeV} & 173\,{\rm MeV} & 1.37\,{\rm GeV} & 4.65\,{\rm GeV}\,,
\end{array}\\ \label{massval}
\end{eqnarray}
and they produce the meson masses in Tables\,\ref{tablevec} and \ref{tableps}.  (NB.\ In solving BSEs to obtain masses the contribution from all orders in the current-quark mass splittings are incorporated.)  We remark that the current-quark masses yield the following dynamical constituent-quark masses via Eq.\,(\ref{MQ}) (in GeV):
\begin{equation}
\begin{array}{lllll}
 M^D_u & M^D_d & M^D_s & M^D_c & M^D_b \\\hline
0.543& 0.548 & 0.669 & 1.662 & 4.771
\end{array}\,. \label{CQmassval}
\end{equation}
For light quarks $M^D_f-m_f = M^D_0-m_f/4$, where $M^D_0$ is the chiral limit value \cite{Bhagwat:2004hn}, and we note that in general $M^D_f-m_f$ is a monotonically decreasing function of $m_f$, bounded below by zero as $m_f\to \infty$.  This result emphasises that the \emph{essentially dynamical} component of chiral symmetry breaking decreases with increasing current-quark mass, as observed previously \cite{Chang:2006bm,Holl:2005st}.

\begin{table}[t]
\caption{\label{tablevec} Vector meson masses calculated from the BSE defined by Eqs.\,(\ref{gendse}), (\ref{Gk}), (\ref{defKA}), (\ref{defxi}) and (\ref{SigmaRL}), using the parameter values in Eqs.\,(\ref{Gval}) and (\ref{massval}).  The experimental values are taken from Ref.\,\cite{Yao:2006px}.  The three parameters and current-quark masses were fitted as described in connection with Eqs.\,(\protect\ref{Gval}) and (\protect\ref{massval}).  See Sec.\,\protect\ref{Nffive} for further discussion of $\rho^0$ and $\omega$.}
\begin{centering}
\begin{tabular}{l|ccc}
 & Expt.\ (GeV) & Calc.\ (GeV) & Th/Ex-1\; (\%)\\\hline
``$\rho^0$'' & 0.7755 & 0.7704 & -0.66\\
$\rho^\pm$& 0.7755 & 0.7755 & 0~~~~\\
``$\omega$'' & 0.7827 & 0.7806 & -0.27 \\
$K^{\ast\pm}$& 0.8917 & 0.8915 & -0.02\\
$K^{\ast 0}$& 0.8960 &  0.8969& ~0.10\\
$\phi$& 1.0195& 1.0195& 0~~~ \\
$D^{\ast 0}$& 2.0067 & 1.8321 & -8.7~~\\
$D^{\ast\pm}$ & 2.0100& 1.8387& ~-8.5~~ \\
$D^{\ast\pm}_s$ & 2.1120 & 1.9871 & ~-5.9~~ \\
$J/\psi$& 3.0969& 3.0969 & 0~~~ \\
$B^{\ast\pm}$&  & 4.8543 & \\
$B^{\ast 0}$&  & 4.8613 &  \\
$B^{\ast 0}_s$&  & 5.0191 & \\
$B_c^{\ast\pm} $& & 6.2047  & \\
$\Upsilon $& 9.4603& 9.4603 & 0~~~ 
\end{tabular}
\end{centering}
\end{table}

The model we're employing is ultraviolet finite and the current-quark masses in Eq.\,(\ref{massval}) cannot be directly compared with QCD's current-quark mass-scales.  Nevertheless, the values are quantitatively consistent with the pattern of flavour-dependence in the explicit chiral symmetry breaking masses of QCD.  It is notable that so far as isospin breaking is concerned, $m_u/m_d=0.52$, which is compatible with other contemporary estimates, e.g., Ref.\,\cite{Yao:2006px}.

Only $\xi$ and $\theta_\xi$ in Eqs.\,(\ref{defKA}) and (\ref{defxi}) remain unknown.  In order to fix these parameters we consider the neutral pseudoscalar mesons.  In this instance Eq.\,(\ref{LadderBSE}) is modified to the extent that $K_L \to K_L + K_A$; i.e., it reads
\begin{eqnarray}
\nonumber
\lefteqn{\Gamma_{H}(k;P)= \int_q
\left[{\cal S}(q_+) \right. }\\
&&  \times \left. \Gamma_{H}(q_+,q_-) {\cal S}(q_-)\right]_{sr} (K_L+K_A)_{tu}^{rs}(q,k;P)\,.
\label{ABSE}
\end{eqnarray}
As noted in connection with Eq.\,(\ref{SigmaRL}), the gap equation is unmodified.  It might be necessary here to emphasise that this single equation describes \emph{all} neutral pseudoscalar bound-states; namely and for example, the only difference between the $\pi^0$, $\eta$ and $\eta^\prime$ is that they are expressed by solutions of the BSE at different values of $P^2$.  From this perspective there is no mixing, as such, but one can pose the question: what is the quark content of each separated state.  As we illustrate in Sec.\,\ref{mixingeep}, the answers can be used to define mixing angles, all of which are in general different.

\begin{table}[t]
\caption{\label{tableps} Pseudoscalar meson masses calculated from the BSE defined by Eqs.\,(\ref{gendse}), (\ref{Gk}), (\ref{defKA}), (\ref{defxi}) and (\ref{SigmaRL}), using the parameter values in Eqs.\,(\ref{Gval}) and (\ref{massval}).  The experimental values are taken from Ref.\,\cite{Yao:2006px}.  The three parameters and current-quark masses were fitted as described in connection with Eqs.\,(\protect\ref{Gval}) and (\protect\ref{massval}).}
\begin{centering}
\begin{tabular}{l|ccc}
 & Expt.\ (GeV) & Calc.\ (GeV) & Th/Ex-1\; (\%)\\\hline
$\pi^0$ & 0.13498 & 0.13460 & ~-0.3\\
$\pi^\pm$& 0.13957 & 0.13499& ~-3.3\\
$K^\pm$& 0.49368 & 0.41703& -15.5\\
$K^0$& 0.49765 &  0.42662& -14.3\\
$\eta$& 0.54751 & 0.45499& -16.9\\
$\eta^\prime$& 0.95778& 0.91960 & ~-4.0\\
$D^0$& 1.8645 & 1.6195 & -13.1\\
$D^\pm$& 1.8693& 1.6270 & -13.0\\
$D^\pm_s$& 1.9682 & 1.7938 & ~-8.9 \\
$\eta_c$& 2.9804& 3.0171 & ~~1.2\\
$B^\pm$& 5.2790 & 4.7747 & ~-9.6\\
$B^0$& 5.2794 & 4.7819 & ~-9.4 \\
$B^0_s$& 5.3675 & 4.9430& ~-7.9\\
$B_c^\pm$& 6.286~~&6.1505  & ~-2.2\\
$\eta_b$& 9.300~ & 9.4438& ~~1.5
\end{tabular}
\end{centering}
\end{table}

The eigenvector for neutral pseudoscalars can be written
\begin{equation}
\label{BSamp0}
\Gamma_{H_5^0}(P) = \sum_{d=0,3,8,\ldots} \, 2\,{\cal F}^a \gamma_5 \left[ ip_1^{a} + \gamma\cdot P p_2^{a}\right],
\end{equation}
with the index selecting all $N_f$ diagonal generators of $U(N_f)$.  Following the steps described above in connection with Eqs.\,(\ref{eigenproblem}) and (\ref{eigenvalueP}), one can obtain the masses and Bethe-Salpeter amplitudes for the neutral bound-states.  At this point we can fix $\xi$ and $\theta_\xi$ through a least-squares fit to the experimental values of $m_{\eta^\prime}$ and $m_{\eta}/m_{\eta^\prime}$.  This procedure yields
\begin{equation}
\label{valxi}
\xi= 0.076\,,\; \theta_\xi = 0\,
\end{equation}
and the masses in Table\,\ref{tableps}.

\subsection{Discussion of Results}
\label{discussion}
\subsubsection{$\eta$--$\eta^\prime$ Mixing}
\label{mixingeep}
We'll begin with this topical issue and at first consider solving the BSE with an eigenvector of the form in Eq.\,(\ref{BSamp0}) but with the sum running only over $d=0,8$; viz,
\begin{equation}
\label{G80}
%
\Gamma_{80}(P) = \, 2\,{\cal F}^8  g^8(P)
+ 2\,{\cal F}^0  g^0(P)\,,
\end{equation}
with $g^j(P)= \gamma_5(ip_1^{j} + \gamma\cdot P p_2^{j})$.  This eigenvector splits off from that associated with ${\cal F}^3$ and the $\pi^0$ in the isospin symmetric limit.  

The case $\xi= 0$ provides a readily understood illustration.  In this case one obtains two bound-state solutions:
\begin{equation}
\begin{array}{ccccc}
{\rm mass\,(GeV)} & p_1^{8} & p_2^{8} & p_1^{0} & p_2^{0} \\
m_{\bar n n} = 0.135 & ~~0.575 & ~~0.047 & 0.814 & 0.067 \\
m_{\bar s s} = 0.622 & - 0.786 & - 0.219 & 0.556 & 0.155
\end{array}\,. \label{ssbarps}
\end{equation}
Focusing on the eigenvectors, we rewrite Eq.\,(\ref{G80}) in the form
\begin{equation}
\Gamma_{80}(0.133) = \cos\theta_1 \, 2\,{\cal F}^8\, \hat g_8^1 - \sin\theta_1 \, 2\,{\cal F}^0 \hat g_0^1\,,\; \theta_1 = -54.7^\circ\,,
\end{equation}
with $\hat g = \gamma_5(i\hat p_1 + \gamma\cdot P \hat p_2)$ where $\hat p_1^2+\hat p_2^2=1$.  This is plainly a solution with ideal-mixing; namely, the lightest solution contains no $s$-quarks and is composed of an equal mixture of $u$- and $d$-quarks.  The Bethe-Salpeter amplitude for the heaviest solution can be written
\begin{equation}
\Gamma_{80}(0.622) = \sin\theta_2 \, 2\,{\cal F}^8\, \hat g_8^2 + \cos\theta_2 \, 2\,{\cal F}^0 \hat g_0^2\,,\; \theta_2 = -54.7^\circ\,,
\end{equation}
which is a pure $\bar s s$ state.  For $\xi=0$ the dynamics decouples from the flavour structure and hence the mixing angles for the two separated states are identical.

With our preferred value of $\xi=0.076$, Eq.\,(\ref{valxi}), we obtain 
\begin{equation}
\begin{array}{ccccc}
{\rm mass\,(GeV)} & p_1^{8} & p_2^{8} & p_1^{0} & p_2^{0} \\
\,m_\eta=0.455 & ~~0.939 & ~~0.219 & 0.250 & 0.090 \\ 
m_{\eta^\prime}=0.924 & - 0.260 &  - 0.077 & 0.876 &  0.400
\end{array}\,,
\end{equation}
from which we infer 
\begin{equation}
\label{valmixing}
\theta_\eta = -15.4^\circ\,,\; 
\theta_{\eta^\prime} = -15.7^\circ\,.
\end{equation}
Thus, while the Dirac structure of the $\eta$ and $\eta^\prime$, described by $g^8$, $g^0$, is different, there is near equality between the mixing angle at each bound-state.  For comparison, from a recent single mixing angle analysis one can extract \cite{KLEO} $\theta = -13.3^\circ \pm 1.0^\circ$.  The angles in Eq.\,(\ref{valmixing}) correspond to the flavour contents:
\begin{eqnarray}
\label{etaetapalone}
|\eta\rangle &\sim& 0.55 \, (\bar u u + \bar d d) - 0.63\, \bar s s \,,\\
|\eta^\prime\rangle &\sim &0.45 \, (\bar u u + \bar d d) + 0.78 \, \bar s s\,.
\end{eqnarray}

\subsubsection{Chiral limit}
In the case of $N_f=3$ massless quarks, Eqs.\,(\ref{gendse}), (\ref{SigmaRL}) and (\ref{ABSE}) produce, without fine tuning, \emph{eight massless} pseudoscalar mesons -- the Goldstone modes -- and \emph{one massive} state.  The massive state is solely associated with ${\cal F}^0$ and 
\begin{equation}
\label{masschiral}
m_{\eta^\prime} \stackrel{{\cal M}=0}{= } 0.852\,{\rm GeV} \,,
\end{equation}
from which follows the model's value of 
\begin{equation}
\frac{\nu_{\eta^\prime}}{f^0_{\eta^\prime}} = (0.770\,{\rm GeV})^2.
\end{equation}
The chiral limit mass in Eq.\,(\ref{masschiral}) is 93\% of the calculated value in Table~\ref{tableps}.

\subsubsection{$\pi^0$--$\eta$--$\eta^\prime$ Mixing}
With three flavours of quark, each with a different mass, all the neutral pseudoscalar mesons ``mix''; i.e., there is no neutral pseudoscalar solution of Eq.\,(\ref{ABSE}) that is associated solely with a single generator of $U(N_f)$.  In this case the eigenvector assumes the form
\begin{equation}
\Gamma_{H_5^0}(P) = \, 2\,{\cal F}^3 g^3(P) + 2\,{\cal F}^8  g^8(P)
+ 2\,{\cal F}^0  g^0(P)
\end{equation} 
and the BSE gives the solutions
\begin{equation}
\begin{array}{lcccccc}
{\rm mass} & p_1^{3} & p_2^{3} & p_1^{8} & p_2^{8} & p_1^{0} & p_2^{0}\\
{\rm (GeV)} & & & & & & \\
0.135 & ~~0.996 & ~~ 0.081 & ~~0.023 & ~~0.002 & 0.009 & 0.001\\
0.455 & -0.026 & -0.006 & ~~0.939 & ~~0.219 & 0.249 & 0.090\\
0.922 & -0.004 & -0.001 & -0.260 & -0.077 & 0.876 & 0.400
\end{array}\,.
\end{equation}
From these Bethe-Salpeter amplitudes one infers the following flavour contents:
\begin{eqnarray}
\label{pi0f}
|\pi^0\rangle & \sim & 0.72 \, \bar u u - 0.69 \, \bar d d - 0.013 \, \bar s s\,, \\
\label{pi8f}
|\eta\rangle & \sim & 0.53\, \bar u u + 0.57 \, \bar d d - 0.63 \, \bar s s\,, \\
\label{pi9f}
|\eta^\prime\rangle & \sim & 0.44\, \bar u u + 0.45 \, \bar d d + 0.78 \, \bar s s \,.
\end{eqnarray}
In the presence of a sensible amount of isospin breaking the $\pi^0$ is still predominantly characterised by ${\cal F}^3$ but there is a small admixture of $\bar ss$.  A glance at Eq.\,(\ref{etaetapalone}) shows that mixing with the $\pi^0$ has a similarly modest impact on the flavour content of the $\eta$ and $\eta^\prime$.  It's effect on their masses is far less.

\subsubsection{$\pi^0$--$\eta$ Mixing}
There is merit in explicating the nature of the flavour-induced difference between the $\pi^0$ and $\pi^\pm$ masses.  If we ignore mixing with mesons containing other than $u,d$-quarks; viz., work solely within $SU(N_f=2)$, then the masses in Eq.\,(\ref{massval}) give $m_{\pi^0}-m_{\pi^+}=-0.04\,$MeV.  On the other hand, it is apparent from Tables~\ref{tablevec} and \ref{tableps} that the full calculation yields $m_{\pi^0}-m_{\pi^+}=-0.4\,$MeV, a factor of ten greater.  When one considers only $SU(N_f=3)$; i.e., a so-called $3$--$8$-mixing, then the $\pi^0$ mass is $0.1\,$MeV larger than in Table~\ref{tableps}: $m_{\pi^0}-m_{\pi^+}=-0.3\,$MeV, and one obtains a mixing angle at the neutral pion mass shell of 
\begin{equation}
\theta_{\pi \eta}(m_{\pi^0}^2)=1.2^\circ.
\end{equation}
For comparison, Ref.\,\cite{Green:2003qw} infers a mixing angle of $0.6^\circ \pm 0.3^\circ$ from a $K$-matrix analysis $p\, d \rightarrow\, ^3$He$\,\pi^0$.  Plainly, mixing with the $\eta$-meson is the dominant non-electromagnetic effect.  Within this subspace, $m_\eta$ is 5\% larger than in Table~\ref{tableps} and 
\begin{equation}
\theta_{\pi \eta}(m_{\eta}^2)=1.3^\circ.
\end{equation}
(Two mixing angles can be introduced to parametrise the complete problem of $\pi^0$--$\eta$--$\eta^\prime$ mixing, e.g., \cite{Leutwyler:1996tz}.
However, that approach contains no information in addition to Eqs.\,(\ref{pi0f}) -- (\ref{pi9f}).)

It is noteworthy that 
\begin{eqnarray}
\frac{\theta_{\pi \eta}(m_{\eta}^2)- \theta_{\pi \eta}(m_{\pi^0}^2)}
{m_{\eta}^2 - m_{\pi^0}^2} & = & r_{\pi \eta}^2\,\theta_{\pi \eta}(m_{\pi^0}^2)\, ,\\
\label{pietaslope}
r_{\pi \eta} & = & 0.582 \, {\rm GeV}^{-1}\,.
\end{eqnarray}
Our DSE framework describes mesons explicitly as bound-states of a dressed-quark and -antiquark.  Hence, it is sensible to compare the result for this mixing angle and its momentum dependence with that, e.g., of Ref.\,\cite{Maltman:1994uu}.  They are commensurate: our mixing angle is $\lesssim 20$\% smaller and the slope in Eq.\,(\ref{pietaslope}) is $\lesssim 20$\% larger.  The slope in Eq.\,(\ref{pietaslope}) is smaller than that which has been calculated in connection with $\rho^0$--$\omega$ mixing but of the same order of magnitude; e.g., Refs.\,\cite{Goldman:1992fi,Krein:1993wv,Mitchell:1994jj}. 

\subsubsection{Five flavours}
\label{Nffive}
The tables present masses calculated with $N_f=5$ flavours of quark.  The vector mesons are least complicated and hence we begin with them.  It is a general feature of the rainbow-ladder truncation that with no two current-quark masses equal, each neutral vector meson is flavour-diagonal; namely, the kernel produces the following states, in order of increasing mass: $\bar u u$, $\bar d d$, $\bar s s$, $\bar c c$, $\bar b b$.  It is a good approximation for the heavier quarks and we therefore used this fact to fix the masses of the $s$-, $c$- and $b$-quarks.  

On the other hand, it is erroneous for the $u,d$-quark neutral vector mesons; viz., experimentally $\rho^0 \neq \bar u u$ and $\omega\neq \bar d d$, and that is why these states are written with quotation marks in Table~\ref{tablevec}.  If one assumes $m_u = m_d$, then there is no discernible problem.  However, in the real case of $m_u \neq m_d$ only an extended kernel can produce the true flavour content for these states.  Improvements along the lines pursued in Refs.\,\cite{Bender:2002as,Bhagwat:2004hn,Matevosyan:2006bk} do not ameliorate the situation because they preserve the flavour structure of the rainbow-ladder truncation.  It is probable that inclusion in the kernel of diagrams which correspond to two- and three-pion intermediate states is required in order to provide the remedy, since the former contribute primarily in the $\bar u u -\bar d d$ channel and the latter predominantly in the $\bar u u +\bar d d$ channel.  Solving the BSE without these channels but with forced $(\bar u u -\bar d d)/\surd 2$ and $(\bar u u +\bar d d)/\surd 2$ flavour contents, the $\rho^0$ and $\omega$ are degenerate with $\rho^\pm$.  An analysis of diagrams corresponding to intermediate states containing pseudoscalar mesons is capable of lifting the $\rho^0$--$\omega$ degeneracy, even for $m_u = m_d$ \cite{Roberts:1988yz,Hollenberg:1992nj,Mitchell:1996dn,Pichowsky:1999mu}. 

Regarding the other entries in Table~\ref{tablevec}, where a determination is possible the model is evidently accurate to $\lesssim 10$\%.  We therefore hold that it is reasonable to expect that the masses predicted for the as yet unobserved $B^\ast$-mesons are similarly accurate.  Indeed, they are probably an underestimate of the physical values by no more than this amount.  

One can read from Table~\ref{tablevec} that for vector mesons composed solely of light-quarks the calculated flavour-dependent mass differences are well approximated by the differences in dynamical constituent-quark masses.  The accuracy is better than 2\%.  We infer from this an estimate of the non-electromagnetic part of the neutron-proton mass difference:
\begin{equation}
[m_n - m_p]_f= 5.3\,{\rm MeV} =: \Delta_{UD}\,.
\end{equation}
This simple projection could be checked via the Faddeev equation approach to nucleon structure \cite{Flambaum:2005kc}, which is kindred to that employed herein for mesons.  For comparison, a numerical simulation of lattice regularised QCD has been used to infer a value for this difference of $2.26 \pm 0.72$ \cite{Beane:2006fk}; and the experimentally determined value, which includes electromagnetic effects, is $1.3\,$MeV.  

Additional context is provided by the observation that this simple reasoning also entails:
$M_{\Sigma^-} - M_{\Sigma^0}= 
M_{\Sigma^0} - M_{\Sigma^+} =
M_{\Xi^-} - M_{\Xi^+} = \Delta_{UD} $.
The experimental values are, respectively: $3.3$, $4.8$, $6.5\,$MeV.  We emphasise that pseudoscalar-meson self-energy diagrams contribute materially to a baryon's mass \cite{hechtfe}.  When dealing with effects of this small magnitude, differences in mass between the mesons that appear in such diagrams will contribute to these mass-differences.  A complete, accurate calculation as opposed to our estimates will naturally require precision.

On the other hand, for vector mesons containing at least one heavy-quark, the difference in current-quark masses provides a better estimate of the non-electromagnetic mass difference.  It is apparent from Table~\ref{tableps} that the same is true of pseudoscalar mesons containing at least one heavy-quark.  

Owing to DCSB, the mass-squared of light-quark pseudoscalar mesons rises linearly with current-quark mass.  This explains our result that for the non-electromagnetic part
\begin{equation}
[m_{K^0} - m_{K^+}]_{f} = 9.6\,{\rm MeV} > [m_{K^{\ast 0}} - m_{K^{\ast +}}]_f = 5.4\,{\rm MeV}\,.
\end{equation}

It is noteworthy that for each system our calculated result for the $H^0-H^+$ mass-difference is larger in magnitude than the experimental difference.  This is a necessary result and a useful check on this and other calculations that omit electromagnetic effects.  The inclusion of electromagnetism will act predominantly to increase the mass of the charged state and hence the mass difference will fall in magnitude.

We have described above the quark-flavour content of neutral pseudoscalar mesons as calculated in subspaces of the full flavour group.  This analysis can be repeated for $U(N_f=5)$ with the resulting flavour probability amplitudes:
\begin{equation}
\begin{array}{l|ccccc}
 & \bar u u  & \bar d d & \bar s s & \bar c c & \bar b b \\\hline
 | \pi^0 \rangle       & 0.72 & -0.69  & -0.014& ~~0.000 & ~~0.0000\\
 | \eta \rangle        & 0.53 & ~~0.57 & -0.63 & -0.022&  -0.0057 \\
 | \eta^\prime \rangle & 0.44 & ~~0.45 & ~~0.78 &  -0.060 &  -0.0141\\
 | \eta_c \rangle      & 0.06 & ~~0.06 & ~~0.05 & ~~0.995 & -0.0037 \\
 | \eta_b \rangle      & 0.02 & ~~0.02 & ~~0.01 & ~~0.005 &  ~~0.9996
\end{array}\,.
\end{equation}
As one might have anticipated, for states not much affected by $K_A$ the probability of finding a particular type of ``hidden flavour'' drops as the mass of the quark flavour increases.  This table indicates that the commonly used approximation of writing neutral light-flavour mesons in the form $c_1 (\bar u u + \bar d d) + c_2 \bar s s $ is accurate at the level of $\lesssim 5$\%.

\begin{figure}[t]
\centerline{
\includegraphics[width=0.48\textwidth]{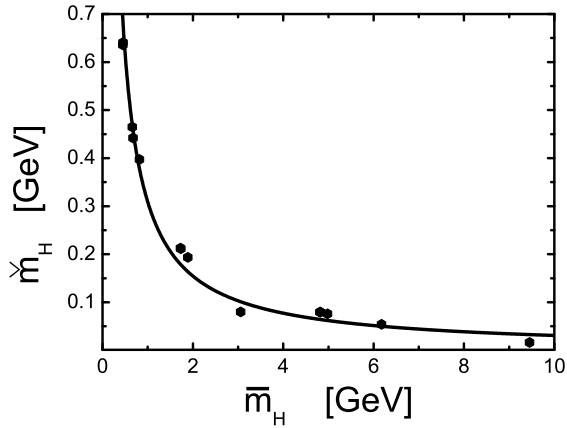}}
\vspace*{2ex}

\caption{\label{VmPsplit} Vector--pseudoscalar meson mass splitting calculated from our results in Tables~\protect\ref{tablevec} and \protect\ref{tableps}; namely, $\check m_H$ as a function of $\bar m_H$, where $\check m_H = m_{H^\ast} - m_{H}$ and $\bar m_H = (1/2)(m_{H^\ast} + m_{H})$.  For the purpose of this figure, we compared $\phi$ with the $\bar s s$ pseudoscalar described in connection with Eq.\,(\ref{ssbarps}).  The solid curve is $\mu/\bar m_H$ with $\mu=0.309\,$GeV$^2$.}
\end{figure}

\subsubsection{Vector--pseudoscalar mass splitting}
As remarked in the Introduction, it is natural to examine the manner by which the $1^-$--$\,0^-$ mass splitting evolves with current-quark mass and mass difference.  It is apparent from Fig.\,\ref{VmPsplit} that without exception the mass splitting, $\check m_H = m_{H^\ast} - m_{H}$, decreases with increasing $\bar m_H = (1/2)(m_{H^\ast} + m_{H})$.  The gross behaviour of the evolution is described well by a $1/\bar m_H$-dependence, which entails
\begin{equation}
\label{massVP}
m_{H^\ast}^2 - m_{H}^2 \sim {\rm const.} = 2\mu = 0.62\,{\rm GeV}^2.
\end{equation}
This outcome is consistent with observation.  It is plain that since $\bar m_H$ is a measure of the dynamical constituent-quark mass, the global picture is not consistent with a simple, single hyperfine interaction between constituent-like quarks.  That is not too surprising owing to the Goldstone boson nature of light pseudoscalar mesons.  Our calculations show that such a picture only becomes reasonable for bound-states containing at least one $c$- or $b$-quark.

\section{Summary}
\label{summary}
In connection with pseudoscalar mesons, the axial-vector Ward-Takahashi identity is a powerful tool whose import should not be ignored.  Following from this and resting upon the empirical observation that the $\eta^\prime$ is not a Goldstone mode, we demonstrated exact chiral-limit relations that connect the dressed-quark propagator to the topological susceptibility.  Furthermore, we extended the mass formulae derived in Refs.\,\cite{Maris:1997hd,Maris:1997tm} to the case of electric-charge-neutral pseudoscalar mesons, for which flavour symmetry breaking entails non-ideal flavour content.  Our development confirms that in the chiral limit the $\eta^\prime$ mass is proportional to the matrix element which connects the $\eta^\prime$ to the vacuum via the topological susceptibility.

To illustrate the implications of the mass formulae we introduced an elementary dynamical model.  This involved an \textit{Ansatz} for that part of the Bethe-Salpeter kernel related to the non-Abelian anomaly which assumes the most general internally consistent form.  It is a key and novel feature of our study that an anomaly contribution is included within the Bethe-Salpeter kernel to yield meson masses.  We thereby avoid the oft used expedient of enforcing anomaly constraints \emph{a posteriori} at the level of a matrix of masses of unphysical-mesons.  

In addition to the current-quark masses our model involves only two parameters, one of which is a mass-scale.  It was employed in a wide-ranging analysis of pseudoscalar- and vector-meson bound-states with an emphasis on the effects of $SU(N_f=2)$ and $SU(N_f=3)$ flavour symmetry breaking.  Section~\ref{discussion} details our findings, which are too numerous to recapitulate here.  Suffice it to report that, despite its simplicity, the model proved elucidative and phenomenologically efficacious.  Our results, both their qualitative and quantitative aspects, should serve as a valuable guide in the future study of neutral pseudoscalar mesons using more realistic interactions.  

\begin{acknowledgments}
This work was supported by: 
Department of Energy, Office of Nuclear Physics, contract no.\ DE-AC02-06CH11357;
the Doctoral Program Foundation of the Ministry of Education, China, under grant No.\ 20040001010;
Major State Basic Research Development Program of China contract no.\
G2007CB815000;
National Natural Science Foundation of China contract nos.\ 10425521, 10575004 and 10675007;
and National Science Foundation grant no.\ PHY-0610129,
One of the authors (Y.-X.~Liu) would also like to acknowledge support from the Foundation for University Key Teacher by the Ministry of Education, China.
\end{acknowledgments}


\begin{thebibliography}{40}
\bibitem{fn:Euclidean} We employ a Euclidean metric, with:  $\{\gamma_\mu,\gamma_\nu\} = 2\delta_{\mu\nu}$; $\gamma_\mu^\dagger = \gamma_\mu$; $ \gamma_5 = - \gamma_1 \gamma_2 \gamma_3 \gamma_4$; ${\rm tr}\, \gamma_5 \gamma_\mu \gamma_\nu \gamma_\rho \gamma_\sigma = -4\, \varepsilon_{\mu\nu\rho\sigma}$;  and $a \cdot b = \sum_{i=1}^4 a_i b_i$.  For a timelike vector $P_\mu$, $P^2<0$.
\bibitem{fn:momentum} We define $k=(1-\eta) p_1 + \eta p_2$, with $\eta \in [0,1]$, and hence $k_+ = p_1 = k + \eta P$, $k_- = p_2 = k - (1-\eta) P$.  In a Poincar\'e covariant approach, such as ours, no observable can depend on $\eta$; i.e., the definition of the relative momentum.  
\end{thebibliography}
\end{document}